\documentclass[twocolumn,groupedaddress,amsmath,amssymb,prl,aps]{revtex4}

\usepackage{graphicx}% Include figure files
\usepackage{dcolumn}% Align table columns on decimal point
\usepackage{bm}% bold math
\usepackage{soul} %strikethrough - MEG
\usepackage{xcolor} % color text - MEG
\usepackage{amsmath} % more tools for math environment - MEG
\usepackage{lipsum}
\usepackage{fdsymbol}
\begin{document}

\title{Brownian electric bubble quasiparticles}

\author{Hugo Aramberri$^1$}
\author{Jorge \'I\~niguez-Gonz\'alez$^{1,2}$}

\affiliation{$^1$Materials Research and Technology Department, Luxembourg Institute of Science and Technology (LIST), Avenue des Hauts-Fourneaux 5, L-4362 Esch/Alzette, Luxembourg}

\affiliation{$^2$Department of Physics and Materials Science, University of Luxembourg, Rue du Brill 41, L-4422 Belvaux, Luxembourg}

\date{\today}

\begin{abstract} Recent works on electric bubbles (including the experimental
  demonstration of electric skyrmions) constitute a breakthrough akin
  to the discovery of magnetic skyrmions some 15 years ago. So far
  research has focused on obtaining and visualizing these objects,
  which often appear to be immobile (pinned) in experiments. Thus,
  critical aspects of magnetic skyrmions -- e.g., their quasiparticle
  nature, Brownian motion -- remain unexplored (unproven) for electric
  bubbles. Here we use predictive atomistic simulations to investigate
  the basic dynamical properties of these objects in pinning-free
  model systems. We show that it is possible to find regimes where the
  electric bubbles can present long lifetimes ($\sim$~ns) despite
  being relatively small (diameter $< 2$~nm). Additionally, we find
  that they can display stochastic dynamics with large and highly
  tunable diffusion constants. We thus establish the quasiparticle
  nature of electric bubbles and put them forward for the physical
  effects and applications (e.g., in token-based Probabilistic
  Computing) considered for magnetic skyrmions.
  \end{abstract}

\maketitle

\begin{figure}
  \includegraphics[width=\columnwidth]{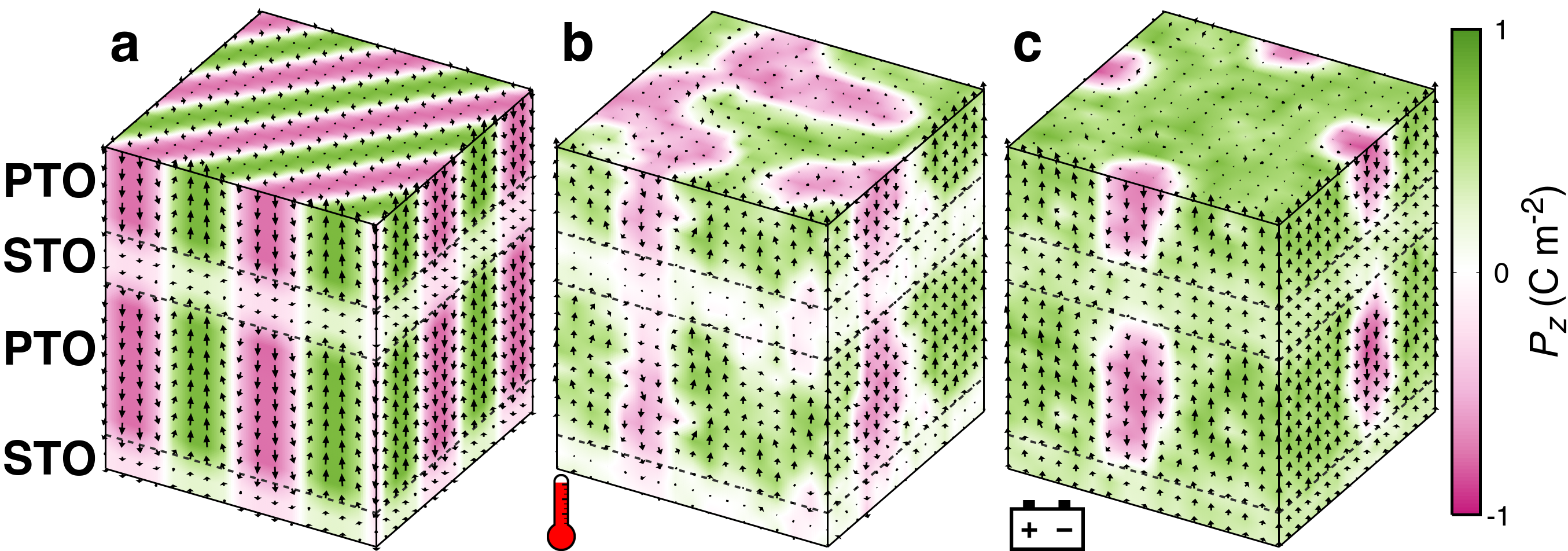}
	\caption{{\bf Domains in
            ferroelectric/paraelectric superlattices.} Stripe domains
          ({\bf a}), ``domain liquid'' state ({\bf b}) with
          spontaneous domain wall motion caused by thermal
          fluctuations, and e-bubbles ({\bf c}) resulting from applying
          an electric field to the stripe phase. Arrows indicate 
          the local polarization in the
          shown planes; the color code indicates the polarization
          along the vertical direction (scale to the
          right).}
	\label{fig:sketch}
\end{figure}

Recent works show that ferroelectrics can present topological textures
akin to magnetic skyrmions~\cite{junquera23,das19,goncalves19}. Many
studies have proven the possibility to stabilize and manipulate
electric bubbles~\cite{lichtensteiger14,zhang17,bakaul21}, sometimes
providing evidence for their non-trivial
topology~\cite{das19,han22}. Perovskites are the best studied
materials so far~\cite{junquera23}, with ferroelectric/paraelectric
PbTiO$_{3}$/SrTiO$_{3}$ superlattices emerging as model systems in the
field (Fig.~\ref{fig:sketch}).

The experimentally observed electric bubbles (``e-bubbles'') seem to
be static, suggesting they are pinned by
defects~\cite{lichtensteiger14,zhang17,bakaul21}. However, many
interesting e-bubble properties will be connected to their potential
particle-like dynamics. This is the case for magnetic
skyrmions~\cite{nagaosa13}, which feature anomalous responses and
Brownian motion~\cite{zazvorka19,nozaki19,jibiki20,raab22}. Skyrmions'
main appeal for applications -- from racetrack
  memories~\cite{kiselev11,fert13,parkin08} to token-based
Probabilistic Computing~\cite{pinna18,jibiki20} -- is their mobile
quasiparticle nature. We believe that e-bubbles can also behave as
mobile quasiparticles.

In 2016 atomistic Monte Carlo simulations~\cite{zubko16} predicted
that the ferroelectric domains of some PbTiO$_{3}$/SrTiO$_{3}$
superlattices present stochastic diffusion in a range of about 100~K
below the Curie point ($T_{\rm C}$). This so-called ``domain liquid''
(Fig.~\ref{fig:sketch}{\bf b} and Supp. Video~1) was supported by
X-ray diffraction experiments showing well-developed local
polarization but no long-range order~\cite{zubko16}. This suggests
that domain walls are not strongly pinned in the studied samples.

The stochastic motion of these domains makes physical sense. In the
investigated ultra-thin layers ($\sim 2$~nm), domain walls have
relatively small areas and the domains themselves are very narrow
($\sim 6$~nm)~\cite{zubko10}; hence, thermal fluctuations (large at
$T\lesssim T_{\rm C}$) may result in net wall shifts, which yield
changes in the size, shape and position of the domains.

\begin{figure*}[t]
  \includegraphics[width=\textwidth]{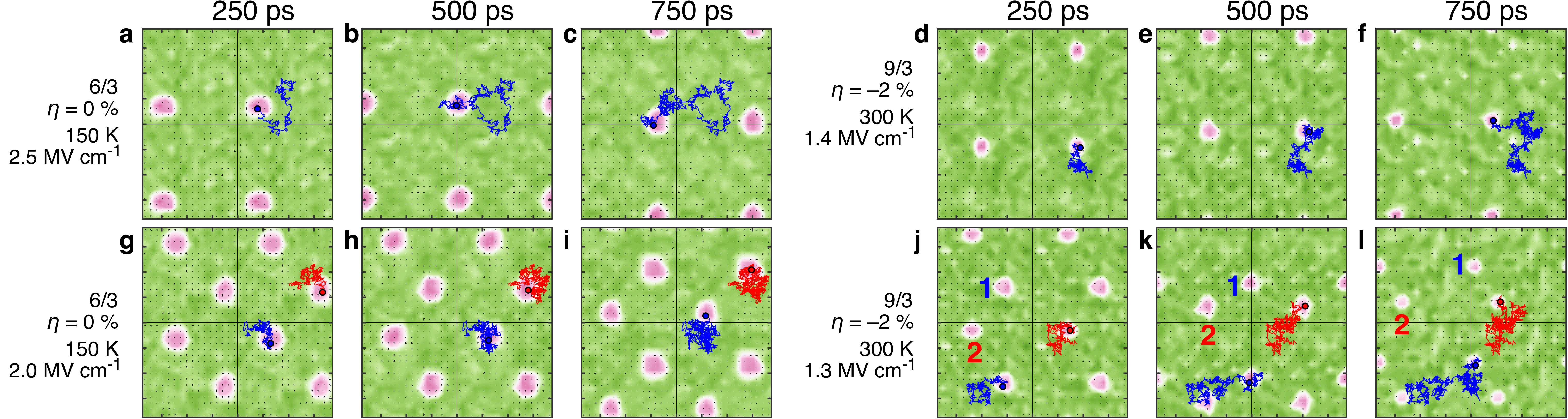}
	\caption{{\bf Stochastic dynamics of the e-bubbles.}
          Snapshots of molecular dynamics runs at the times indicated
          above. The first (second) row shows cases with one (two)
          e-bubble(s) per simulation supercell. We show the dipoles at
          the middle of the PbTiO$_{3}$ layer. We display four
          periodic images of the simulation supercell. The in-plane
          simulation supercell ($16\times 16$ elemental perovskite
          units) is marked by black lines. {\bf a}-{\bf c} and {\bf
            g}-{\bf i} correspond to a 6/3 superlattice at $\eta =
          0$~\% and $T = 150$~K, under ${\cal E}=2.5$~MV~cm$^{-1}$ and
          2.0~MV~cm$^{-1}$, respectively. {\bf d}-{\bf f} and {\bf
            j}-{\bf l} correspond to a 9/3 superlattice at $\eta =
          -2$~\% and $T = 300$~K, under ${\cal E}=1.4$~MV~cm$^{-1}$
          and 1.3~MV~cm$^{-1}$, respectively. The color scale is that
          of Fig.~\ref{fig:sketch}.}
	\label{fig:path}
\end{figure*}

Critically, several groups have shown that stripe domains
(Fig.~\ref{fig:sketch}{\bf a}) can be broken into e-bubbles
(Fig.~\ref{fig:sketch}{\bf c}, Supp. Video~2) under an electric field
${\cal E}$~\cite{kornev04,govinden23,murillo-navarro23}. This suggests
that applying an electric field to the ``domain liquid'' state should
result in Brownian e-bubbles. Here we explore this possibility using
the atomistic simulation methods~\cite{wojdel13} that predicted
electric skyrmion bubbles (experimentally
verified~\cite{das19,goncalves19}) and ferroelectric domain liquids
(experimentally supported~\cite{zubko16,gomez-ortiz23b}) in these
systems. Our results reveal strategies to stabilize Brownian e-bubbles
with tunable properties (density, lifetime, diffusion speed) and
establish their nature as mobile quasiparticles.

We build on recent investigations of PbTiO$_{3}$/SrTiO$_{3}$
superlattices under electric
fields~\cite{murillo-navarro23,aramberri22,graf22}, considering
substrates like SrTiO$_{3}$ or slightly more compressive, so the polar
axis of PbTiO$_{3}$ is along the stacking direction. The zero-field
ground state is characterized by stripe ferroelectric domains that
form to minimize depolarizing fields (Fig.~\ref{fig:sketch}{\bf a}).

If we apply an electric field that favors the upward polarization, the
PbTiO$_{3}$ layers typically react in two stages. For small fields the
domain walls get polarized, which usually yields a
negative-capacitance response~\cite{zubko16,graf22,iniguez19}. With
increasing field, the downward-polarized stripes eventually break,
yielding disconnected downward domains within an upward-polarized
matrix (Fig.~\ref{fig:sketch}{\bf
  c})~\cite{kornev04,govinden23,murillo-navarro23}. These minority
domains are our e-bubbles. Such a strategy to stabilize e-bubbles has
been discussed previously~\cite{kornev04,govinden23} and mimics
standard ways to obtain skyrmions from spin spirals by applying
magnetic fields~\cite{nagaosa13}.

We explore the phase diagram of these superlattices as a function of
temperature, electric field and their basic defining parameters
(epitaxial strain $\eta$ and thickness of the PbTiO$_{3}$
layer). Supp. Note~1 and Supp. Figs.~S1-S4 summarize our results,
which lead to the following main observations.

At low temperatures the e-bubbles are immobile. They start to diffuse
upon moderate heating, up until they dissolve into a strongly
fluctuating matrix at higher temperatures. This sequence mimics the
one observed at zero bias: frozen stripe domains at low temperatures,
then a ``domain liquid'' and finally a paraelectric state at high
temperature~\cite{zubko16,gomez-ortiz23b}. The stability range of the
``e-bubble liquid'' can be pushed to higher temperatures by
strengthening the polar order (e.g. with compressive epitaxial strains
or thicker PbTiO$_{3}$
layers~\cite{aramberri22,graf22,murillo-navarro23}).

Upon varying the applied field ${\cal E}$, the PbTiO$_{3}$ layers
adapt mainly by adjusting the number of
e-bubbles~\cite{govinden23,nahas20}. By contrast, the in-plane bubble
diameter stays roughly constant; it seems to be determined by the
PbTiO$_{3}$ layer's thickness in the same way as Kittel's law controls
the width of the stripe domains at zero bias~\cite{gomez-ortiz23a}.

The e-bubbles can be longevous. Typically, they remain distinct
particles through the whole molecular dynamics runs (up to 3~ns),
despite their evolving shape and position. Nevertheless, we also find
conditions where the e-bubbles (dis)appear spontaneously. This seems
to happen in regimes where states with different e-bubble density
compete. Also, e-bubbles are shorter lived at higher temperatures.

Now we focus on the dynamics of individual e-bubbles. We first
consider conditions yielding single long-lived bubbles in our
simulation supercell. Let us take the representative case of a 6/3
superlattice (with PbTiO$_{3}$ and SrTiO$_{3}$ layers that are 6- and
3-unit-cells thick, respectively) at $\eta = 0$~\%, ${\cal E} =
2.5$~MV~cm$^{-1}$ and 150~K. Figs.~\ref{fig:path}{\bf a}-{\bf c} show
the diffusion of the e-bubble (see Suppl. Video~3). The jerky
trajectory suggests a stochastic motion, resembling recent
experimental results for Brownian magnetic
skyrmions~\cite{zazvorka19,nozaki19,raab22}, except here at shorter
length and time scales (nm instead of $\mu$m, ps instead of ms).

\begin{figure}
  \includegraphics[width=\columnwidth]{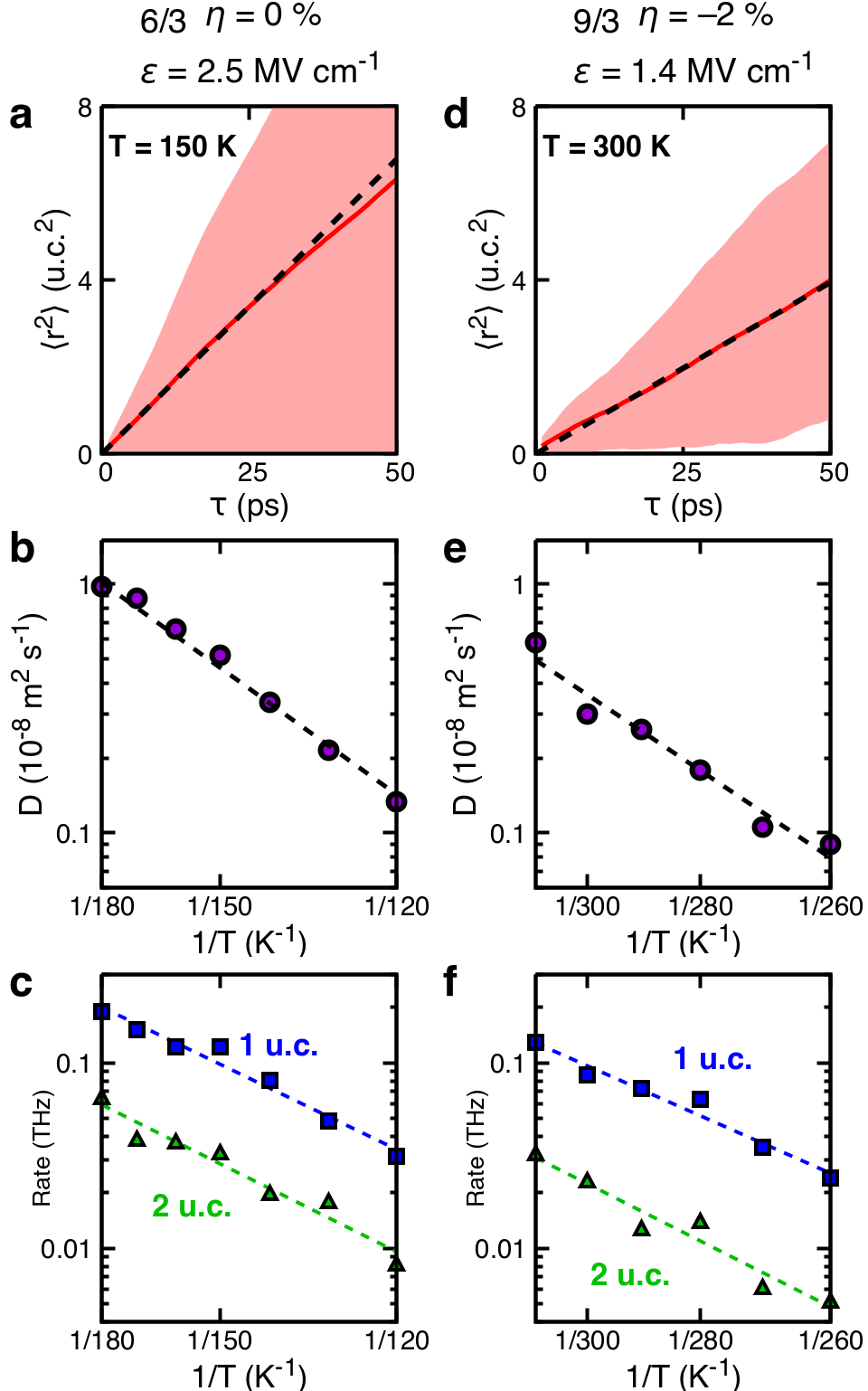}
	\caption{{\bf Activated diffusion of e-bubbles}. Left: 6/3 superlattice at
          $\eta = 0$~\% and ${\cal E} = 2.5$~MV~cm$^{-1}$ ({\bf a}-{\bf c}). Right: 9/3
          system at $\eta = -2$~\% and ${\cal E} = 1.4$~MV~cm$^{-1}$
          ({\bf d}-{\bf f}). Mean squared
          displacement $\langle r^2 \rangle$ as a function of time (obtained at 150~K in
          {\bf a}, at 300~K in {\bf d}), diffusion coefficient ({\bf b} and {\bf e}) and hopping rates ({\bf c} and {\bf f}) as a function of inverse temperature. In {\bf c} and
          {\bf f} we show rates corresponding to jumps by 1
          (blue) and 2 (green) unit cells. Dashed
          lines in {\bf b}-{\bf f} indicate
          Arrhenius fits. Shaded areas ({\bf a} and {\bf d}) indicate the variance of $r^2$.}
	\label{fig:graphs}
\end{figure}

Figure~\ref{fig:graphs}{\bf a} shows the mean squared displacement
$\langle r^{2} \rangle$ of this trajectory. Up to 25~ps we observe a
linear time dependence. We can thus assume the form
\begin{equation}
  \langle r^{2} \rangle = 2 {\rm d} D t + {\cal O}(t^{2}) \; ,
\end{equation}
where $D$ is the diffusion constant, ${\rm d}$ is the dimension of the
space where the e-bubble moves (here ${\rm d}=2$), and nonlinear terms
are corrections to an ideal random walk. For this example we obtain $D
= 5.2\cdot 10^{-9}$~m$^2$s$^{-1}$.

As shown in Fig.~\ref{fig:graphs}{\bf b}, $D$ depends exponentially on
the inverse temperature. This is typical of thermally-activated
diffusion in solids~\cite{mehrer-book07}. Figure~\ref{fig:graphs}{\bf
  b} shows the fit to an Arrhenius law
\begin{equation}
  D = D_{0} \exp{(-E_{a}/k_B T)} \; , 
\end{equation}
where $k_{\rm B}$ is Boltzmann's constant, $E_{a} = 61 \pm 3$~meV is
the activation energy and $D_{0} = (51 \pm 3) \cdot
10^{-8}$~m$^{2}$s$^{-1}$. We also collect statistics for the hopping
rate of the e-bubbles and study its temperature dependence. As shown
in Fig.~\ref{fig:graphs}{\bf c}, the average rate presents an
Arrhenius behavior, the details depending on the hopping distance used
in the analysis: we get $E_{a} = 55 \pm 5$~meV if one unit cell (u.c.)
is considered, and $E_{a} = 62 \pm 6$~meV for jumps of 2~u.c.

To gain insight into these activation energies, let us recall the
literature on domain walls in bulk PbTiO$_{3}$. According to
first-principles studies~\cite{meyer02,wang14}, the rigid shift of a
180$^{\circ}$ wall has to overcome an energy barrier of
$\approx$37~mJ~m$^{-2}$. Noting that our e-bubbles are essentially
wrapped by a 180$^{\circ}$ wall, and that this wall has an area of
about 9.5~nm$^{2}$ for the bubbles in the 6/3 superlattice, the
results of Refs.~\onlinecite{meyer02} and \onlinecite{wang14} suggest
an activation energy of $\approx$2.2~eV. Interestingly, if we simulate
the rigid shift of the e-bubble, we obtain a barrier of similar
magnitude: 1.2~eV (Supp. Fig.~S5).

In reality walls do not shift rigidly, but deform to follow
lower-energy paths. Using accurate atomistic potentials, Shin {\sl et
  al}.~\cite{shin07} obtained a dominant activation energy of 100~meV
for the (field-driven) motion of walls in bulk PbTiO$_{3}$ between
200~K and 300~K. Such a barrier is close to the values around 60~meV
we get for the diffusion of e-bubbles. This in-magnitude agreement
supports the physical soundness of our results.

We also predict Brownian e-bubbles at room temperature in
superlattices with stronger polar order. Figures~\ref{fig:path}{\bf d}
to \ref{fig:path}{\bf f} show representative results for a 9/3 system
with $\eta = -2$~\% and ${\cal E}=1.4$~MV~cm$^{-1}$. Here we observe
frequent events where an e-bubble (dis)appears, with an estimated
average lifetime of hundreds of picoseconds (see Supp. Video~4.). We
get $D = 3.0 \cdot 10^{-9}$ m$^2$s$^{-1}$ at 300~K and a
thermally-activated hopping with $E_{\rm a} = 254 \pm 26$~meV (see
Figs.~\ref{fig:graphs}{\bf d}-{\bf f}). This activation energy is
about 4~times larger than that mentioned above for the 6/3
superlattice with $\eta = 0$~\%. (As shown in Supp.  Fig.~S5, we
obtain a similar increase in the barrier associated to the rigid shift
of a bubble.) This suggests a strong dependence of $E_{\rm a}$ on the
area of wall wrapping the bubble.

What happens when the e-bubbles are not isolated?
Figures~\ref{fig:path}{\bf g}-{\bf i} show representative results of
two bubbles evolving at 150~K in the aforementioned 6/3 superlattice,
here subject to ${\cal E} = 2.0$~MV~cm$^{-1}$ (see
Supp. Video~5). Despite their proximity, the bubbles remain distinct
particles. Yet, they are ordered and form a (drifting) lattice, with a
smaller $D = 3.9 \cdot 10^{-9}$~m$^{2}$s$^{-1}$. By contrast,
Figs.~\ref{fig:path}{\bf j}-{\bf l} show two e-bubbles at 300~K in the
aforementioned 9/3 superlattice, here under ${\cal E} =
1.3$~MV~cm$^{-1}$ (see Supp. Video~6). These bubbles diffuse with $D =
3.7\cdot 10^{-9}$~m$^{2}$s$^{-1}$, exceeding the above mentioned
results for isolated particles in essentially the same system ($D =
3.0\cdot 10^{-9}$~m$^{2}$s$^{-1}$ for ${\cal
  E}=1.4$~MV~cm$^{-1}$). This surprising result is consistent with the
energy barriers we obtain in the limit of rigid bubble shifts
(Supp. Fig.~S5), suggesting a direct impact of the field in the
hopping rates.

The bubbles interact. Inspection of the two-bubble trajectories of
Fig.~\ref{fig:path} shows that their centers never get closer than
2.5~nm, evidencing a short-range repulsion. Further, the bubbles of
Figs.~\ref{fig:path}{\bf j}-{\bf l} show signs of mid-range
attraction. For example, bubbles ``1'' and ``2'' in
Fig.~\ref{fig:path}{\bf j} are only about 3.5~nm away, and continue
linked for at least 250~ps (Fig.~\ref{fig:path}{\bf k}). Eventually
that link breaks and new links form (Fig.~\ref{fig:path}{\bf l}). Due
to the finite size of our simulation supercell, we cannot resolve the
details of this attractive interaction. Yet, our results do suggest
the bubbles tend to approach (transitorily) with a high
probability. (An attractive interaction can be deduced also from the
trajectory depicted in Figs.~\ref{fig:path}{\bf g}-{\bf i}, see
Supp. Note~2 and Supp. Fig.~S6.)

\begin{figure}
  \includegraphics[width=\columnwidth]{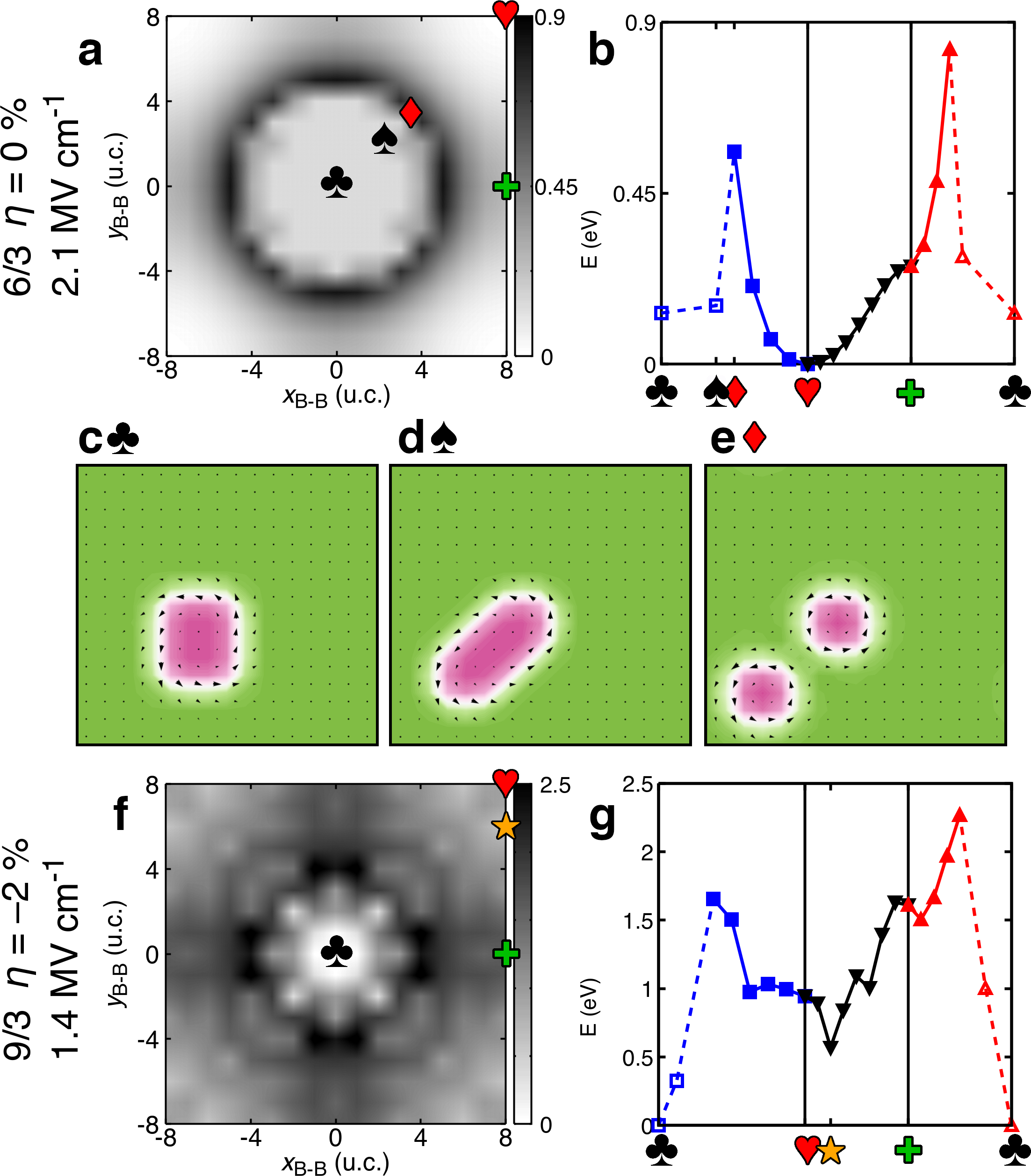}
	\caption{{\bf Inter-bubble interactions.} {\bf a} Computed
          energy map of a two-bubble system, with one bubble at the
          center and a second at every other possible position of the
          simulation supercell. {\bf b} Energy profiles between points
          marked with symbols in {\bf a}. {\bf c}-{\bf e}
          Configurations marked with the corresponding symbols in {\bf
            a}. The central low-energy region inside the dark ring in
          {\bf a} corresponds to merged e-bubbles. We use a 6/3
          superlattice with $\eta = 0$~\% and ${\cal E} =
          2.1$~MV~cm$^{-1}$, nominally at 0~K. The in-plane electric
          dipoles at the bubble walls confer them a Bloch-type
          skyrmion character (see Supplementary Note~4
            for more). {\bf f}-{\bf g} show similar results for a 9/3
          superlattice at $\eta = -2$~\% under ${\cal E} =
          1.4$~MV~cm$^{-1}$.}
	\label{fig:pots}
\end{figure}

To better understand inter-bubble couplings, we compute the potential
energy of two particles at different relative
positions. Figures~\ref{fig:pots}{\bf a}-{\bf e} show results for the
6/3 superlattice with $\eta = 0$~\% and ${\cal E} =
2.1$~MV~cm$^{-1}$. We obtain a nearly central potential with a
lowest-energy region for inter-bubble distances of about 4.4~nm
(\textcolor{red}{$\varheartsuit$} in Fig.~\ref{fig:pots}{\bf a}) and a
central well corresponding to states with only one big bubble
($\clubsuit$ in Fig.~\ref{fig:pots}{\bf a}). These two stable
configurations differ by 134~meV and are separated by a barrier of at
least 0.56~eV. This is consistent with the fact that the e-bubbles in
Figs.~\ref{fig:path}{\bf g}-{\bf i} tend to occupy the whole
simulation supercell.

Figures~\ref{fig:pots}{\bf f} and \ref{fig:pots}{\bf g} show the
potential corresponding to the 9/3 superlattice with $\eta = -2$~\%
and ${\cal E}= 1.4$~MV~cm$^{-1}$. Here we also observe a barrier
separating the one- and two-bubble states. Remarkably, here the most
favorable two-bubble configuration (orange star in
Figs.~\ref{fig:pots}{\bf f}-{\bf g}) does not correspond to the
maximum inter-bubble distance allowed by our supercell. This implies
attractive forces, consistent with the formation of transient pairs in
Figs.~\ref{fig:path}{\bf j}-{\bf l}. (See Supp. Figure~S7.)

We can make a few observations about the origin of the inter-bubble
interactions. The central energy well in Figs.~\ref{fig:pots}{\bf a}
and \ref{fig:pots}{\bf f} is favored by the reduction in domain wall
area corresponding to having one big bubble instead of two small ones
(with similar total in-plane section). By contrast, electrostatics
favors separated bubbles: this reduces depolarizing fields and enables
well-developed local polar order in PbTiO$_{3}$. The high energy
barrier corresponds to configurations with two distinct close
e-bubbles (large wall area, inefficient electrostatic
screening). Finally, the long-range tails and the origin of the
attractive interaction are hard to address. Resolving this will
require calculations in bigger supercells with isolated bubble
pairs. It remains for future work.

From the potential of Fig.~\ref{fig:pots}{\bf a}, we can quantify the
spring constant of bubble pairs. We obtain $\kappa \approx 0.8 \cdot
10^{-3}$~eV\AA$^{-2}$. As explained in Supp. Note~2 and Supp. Fig.~S6,
we can process the simulations (for the representative superlattice at
150~K) to quantify the frequency with which the inter-bubble distance
changes by 1 u.c., obtaining $\nu = 0.06$~THz. This suggests that the
bubbles have an effective mass $m = 2 \kappa / (2\pi\nu)^2 = 1.8 \cdot
10^{-25}$~kg $\approx 0.5 m_{\rm Pb}$, where $m_{\rm Pb}$ is the mass
of a Pb atom. We can estimate the bubble mass in other ways, obtaining
values between $7.6\cdot10^{-25}$~kg (based on an effective spring
constant at 150~K, see Supp. Note~2 and Supp. Fig.~S6) and $1.8 \cdot
10^{-27}$~kg (from the average squared velocity of the e-bubble, see
Supp. Note~3 and Supp. Fig.~S8). This spread of values reflects the
challenges to model the dynamics of e-bubbles.

To conclude, let us contextualize our predictions and stress their
implications. First, note that the simulated e-bubbles are very fast,
with diffusion constants of $D = 3.7 \cdot 10^{-9}$~m$^{2}$s$^{-1}$ at
room temperature, while the fastest Brownian magnetic skyrmions
reported range from $D = 10^{-11}$~m$^{2}$s$^{-1}$
(experiment~\cite{zazvorka19,jibiki20}) to $D =
10^{-8}$~m$^{2}$s$^{-1}$ (simulations~\cite{zazvorka19}). Future work
on the factors controlling e-bubble diffusion will undoubtedly uncover
strategies to speed them up.

Our results suggest that e-bubbles offer an alternative implementation
of concepts proposed for Brownian magnetic skyrmions. An example is
token-based Probabilistic Computing, where it was shown that the
stochastic motion of the skyrmions can power energy-efficient signal
reshufflers and other devices~\cite{pinna18,zazvorka19,jibiki20}. In
Reservoir and Neuromorphic Computing, the nonlinear response of
Brownian skyrmions was leveraged to create low-power logic
gates~\cite{raab22} or neural networks~\cite{yokouchi22}. Given the
distinctive features of the e-bubbles (fast, small, controllable with
electric fields), they might prove useful in such applications, and
even become the quasiparticle of choice for further developments
(e.g. miniaturization and speedup).

Having predicted the mobility and quasiparticle nature
  of the e-bubbles, the next natural goal is to devise a way to
  transport them by application of external fields, in analogy to what
  can be done with magnetic skyrmions~\cite{nagaosa13,raimondo22}. We
  think we can achieve this through electric field or temperature
  gradients, which will allow us to impose a spatial variation of the
  e-bubble density, yielding a net bubble
  current in turn. Exploring this possibility is our current focus.

Pinning will be a difficulty for the experimental demonstration of
Brownian e-bubbles. It will result in a reduced number of mobile
particles and a slower diffusion. Yet, there are strategies to favor
depinning (working close to $T_{\rm C}$, optimizing the bubble size,
training the samples). Note also that low-frequency operation is
preferred for some applications~\cite{lee23}. Hence, we think that
ultimately pinning will not be critical.

Besides these applied prospects, our results are a doorway to
mesmerizing fundamental questions. For example, for smaller fields we
find regimes where mobile stripes coexist with e-bubbles, merging and
splitting stochastically. Such highly-tunable states may offer a
convenient sandbox to study the role of criticality (within the
so-called ``critical brain hypothesis''~\cite{chialvo10,hesse14}) in
Neuromorphic Computing applications. Even isolated e-bubbles are
fascinating. We are particularly drawn by their rich internal dynamics
that could lead to non-trivial behaviors, potentially active
matter-like~\cite{romanczuk12}. Other interesting
  aspects (e.g., the nature of the electric dipoles, the material
  requirements to obtain e-bubbles) are briefly addressed in
  Supplementary Note~4.

With many open questions and exciting possibilities ahead, we expect
e-bubbles will become a focus of attention. We hope this work will
steer research efforts towards demonstrating, studying and exploiting
the quasiparticle nature of e-bubbles. We think this will constitute the
coming of age of the field, closing the gap with the work on magnetic
skyrmions and hopefully enabling relevant interdisciplinary research,
e.g. in Unconventional Computing.

\begin{acknowledgments}
We are thankful for inspiring discussions with the members of the
TOPOCOM Marie Sk{\l}odowska-Curie Doctoral Network, particularly
D.R.~Rodrigues (Bari), S.~Komineas (Crete) and M.~Kl{\"a}ui
(Mainz). Work funded by the Luxembourg National Research Fund through
Grant C21/MS/15799044/FERRODYNAMICS.
\end{acknowledgments}

\appendix

\section{Appendix on simulation methods}

We simulate PbTiO$_{3}$/SrTiO$_{3}$ using the so-called
second-principles approach as implemented in the \textsc{SCALE-UP}
package~\cite{wojdel13,garciafernandez16,escorihuelasayalero17}. The
models for the superlattices were derived from models for bulk
PbTiO$_{3}$ and SrTiO$_{3}$ used in previous
works~\cite{wojdel13,wojdel14a}. Details of the superlattice models
can be found in Refs.~\onlinecite{zubko16} and
\onlinecite{aramberri22}.

We use periodic boundary conditions and simulation boxes of
16$\times$16 elementary perovskite unit cells in the plane, to allow
for bubbles in relatively dilute regimes. The simulation box contains
only one repetition of the superlattice along the stacking
direction. We take the an SrTiO$_{3}$ substrate as the zero of
epitaxial strain, $\eta = 0$.

For each temperature, epitaxial strain and field considered, we first
run Monte Carlo simulations for a total of 50,000 sweeps. We always
take the ground state at zero applied field as the initial
configuration (obtained via a previous Monte Carlo simulated
annealing), an example of which is shown in Fig.~\ref{fig:sketch}{\bf
  a}. The first 10,000 sweeps are considered as thermalization. From
the remaining 40,000 sweeps we obtain the average equilibrium strains.

We then run a velocity-rescaling molecular dynamics simulation to
obtain thermalized velocities. To this end, we take the final atomic
structure obtained with Monte Carlo as the initial configuration, and
we fix the strain to the equilibrium values previously computed. We
run this simulation for a total of 20~ps with a time step of
0.5~fs. We remove the velocity of the center of mass and we rescale
the velocities only every 100 steps (50~fs) to avoid the ``flying
ice-cube effect''~\cite{harvey98}.

Starting from the final configuration of the velocity-rescaling run,
we then perform a microcanonical (NVE) molecular dynamics simulation
using the velocity Verlet algorithm with a time step of 0.5~fs and for
a total of at least 750~ps. This NVE simulation gives us the data for
the statistical analysis described in this work.

In order to track the bubble center ${\bf{r}_{\rm B}}$ -- as seen
e.g. in the trajectories in Fig.~\ref{fig:path} -- we proceed as
follows. We first compute the electric dipoles in each cell of the
PbTiO$_{3}$ layer, centered at the Pb atoms. To do that, we adopt the
usual linear approximation where the dipoles are obtained from the
product of the atomic Born charges and the atomic displacements with
respect to a high-symmetry (cubic-like) reference structure (see
Refs.~\onlinecite{zubko16} and \onlinecite{aramberri22} for
details). Then, for simplicity, we focus on the central plane of the
PbTiO$_{3}$ layer and derive the corresponding 2D polarization field
${\bf P}(i)$, where $i$ labels cells in the plane. Next we find the
local minima of $P_z(i)$, which allows us to identify the rough
location of the e-bubbles (potentially more than one) in the
simulation supercell. Then, given a local minimum of $P_z(i)$ at cell
$i$, we compute the position of the corresponding e-bubble as
\begin{equation}
    {\bf{r}}_{\rm B}=\frac{ \sum_{j \in \Omega(i)} P_z(j) {\bf r}(j)} { \sum_{j \in \Omega(i)} P_z(j) } \; ,
\end{equation}
where ${\bf r}(j)$ is a 2D vector giving the position of cell $j$ and
$\Omega(i)$ is the set of cells within a radius of 3~unit cells (u.c.)
of cell $i$ and belonging to the considered plane. Hence, in essence,
the e-bubble center corresponds to the ``center of mass'' of the
corresponding region of negative polarization.

To obtain the mean squared displacements shown in
Fig.~\ref{fig:graphs} we proceed as follows. We compute a histogram of
the distance traveled by the bubble after a time $t$ using a bin size
of 0.02~u.c. The mean squared displacement at that $t$ is obtained as
the second moment of the resulting histogram~\cite{menssen19}. This
process is repeated for times up to 50~ps in steps of 0.5~ps.

To obtain the energy maps described in Fig.~\ref{fig:pots} we
initialize the simulation with two bubbles, each with an area of
2$\times$2 unit cells, at different relative positions within a
simulation box with 16$\times$16 in-plane cells. We then relax the
structure by performing a Monte Carlo simulated annealing, starting
from 10~K with an annealing rate of 0.9975 for 10,000 sweeps.

Finally, let us note that here we work at relatively
  high temperatures (i.e., between 120~K and 300~K), so we can get
  statistically meaningful results for the diffusion constants from
  relatively short (computationally feasible) NVE simulations.

For additional methodological considerations, see
  Supplementary Note~5.

\end{document}